\documentstyle[11pt]{article}
\begin{document}
\vspace*{.7cm}
\begin{flushright}
\large{UND-HEP-94-BIG09\\
HUTP-94/A029\\
October 1994\\
}
\end{flushright}
\vspace{1.2cm}
\begin{center}
\LARGE{\bf INTERFERENCE BETWEEN CABIBBO ALLOWED
AND DOUBLY FORBIDDEN TRANSITIONS IN
$D\rightarrow K_{S,L}+ \pi 's$ DECAYS}
\end{center}

\vspace{1cm}
\begin{center}\Large
I.I. Bigi\\
\vspace*{.4cm}
{\normalsize
{\it Dept. of Physics, University of Notre Dame du Lac,
Notre Dame, IN 46556, U.S.A.} \\
{\it e-mail address: BIGI@UNDHEP.HEP.ND.EDU}}\\
and \\
H. Yamamoto\\
\vspace*{.4cm}
{\normalsize
{\it High-Energy Physics Lab.,
Harvard University, 42 Oxford St., Cambridge,
MA 02138, U.S.A.}\\
{\it e-mail: YAMAMOTO@HUHEPL.HARVARD.EDU}}

\vspace{.4cm}
\end{center}

\thispagestyle{empty}\vspace{.4cm}
\centerline{\Large \bf Abstract}

Both Cabibbo allowed and doubly forbidden transitions contribute
coherently to $D\rightarrow K_{S,L}+\pi 's$ decays. This leads to several
intriguing and even quantitatively significant consequences, among them:
(i) A difference between $\Gamma (D^+\rightarrow K_S \pi ^+)$ and
$\Gamma (D^+\rightarrow K_L \pi ^+)$ and between
$\Gamma (D^0\rightarrow K_S \pi ^0)$ and
$\Gamma (D^0\rightarrow K_L \pi ^0)$ of roughly 10\% ; similarly
$\Gamma (D^+\rightarrow [K_S\pi ^0]_{K^*} \pi ^+)
\neq \frac{1}{4}\Gamma (D^+\rightarrow [K^-\pi ^+]_{K^*} \pi ^+)$,
and more generally $\Gamma (D\rightarrow \bar K^0+\pi 's) \neq
2\Gamma (D\rightarrow K_S+\pi 's)$.
(ii) A change in the relative phase between the isospin 3/2 and 1/2
amplitudes as extracted from the observed branching ratios for
$D^+\rightarrow K_S\pi ^+$, $D^0\rightarrow K_S\pi ^0 ,\, K^-\pi ^+$.
(iii) If New Physics intervenes to provide the required {\em weak}
phase, then CP asymmetries of up to a few per cent can arise in
$D^+\rightarrow K_S\pi ^+$ vs. $D^-\rightarrow K_S\pi ^-$,
$D^0\rightarrow K_S\pi ^0$ vs. $\bar D^0\rightarrow K_S\pi ^0$,
$D^+\rightarrow [K_S\pi ^0]_{K^*}\pi ^+$ vs.
$D^-\rightarrow [K_S\pi ^0]_{K^*}\pi ^-$, etc.;
an asymmetry of the same size, but opposite in sign occurs when the
$K_S$ is replaced by a $K_L$ in the final state.

\newpage
\large
\addtocounter{page}{-1}

The weak decays of charm hadrons possess three layers, namely
Cabibbo allowed, once and twice forbidden transitions. The first
two layers have clearly been observed in $D$ decays, while the third
one is now emerging in the data.

In this note we want to point out that the existence of doubly
Cabibbo suppressed $D$ (DCSD) channels has a subtle, yet significant
impact on Cabibbo allowed $D$ (CAD) decays producing neutral kaons. The
main results are:

\noindent $\bullet$ $\Gamma (D\rightarrow K_S + \pi 's) \neq
\Gamma (D\rightarrow K_L + \pi 's)$ (irrespective of CP violation in
$K\rightarrow \pi \pi$ decays) and thus
$\Gamma (D\rightarrow \bar K^0+\pi 's)\neq 2\Gamma (D\rightarrow K_S+\pi 's)$;
the difference allows to extract the
amplitude for the DCSD mode $D\rightarrow K^0 + \pi 's$.

\noindent $\bullet$ Ignoring DCSD leads to a systematic
error of ${\cal O}(10\% )$ in the
extraction of the final state phaseshifts in the $D\rightarrow K\pi$ etc.
channels.

\noindent $\bullet$ If New Physics intervenes in DCSD and provides a weak
phase,
then $D\rightarrow K_S + \pi 's$ decays would exhibit a direct CP asymmetry
that (i) cancels against the corresponding asymmetry in
$D\rightarrow K_L + \pi 's$ vs. $\bar D\rightarrow K_L + \pi 's$ modes,
(ii) establishes the existence of physics beyond the KM ansatz for
CP violation and (iii)
can be interpreted in terms of the microscopic parameters of the New Physics.

\noindent These three topics will be discussed in the following.

\section{$D\rightarrow K_S+\pi 's$ vs.
$D\rightarrow K_L+\pi 's$ Decays}

\subsection{$D\rightarrow K_{S,L}\pi $}

The final state $K_S \pi ^+$ can be reached through
$D^+\rightarrow \bar K^0 \pi ^+$ as well as through
$D^+\rightarrow K^0 \pi ^+$. Accordingly one finds for the transition
amplitude a coherent superposition
$$T(D^+\rightarrow K_S \pi ^+)= \frac{1}{\sqrt{2}}
T(D^+\rightarrow \bar K^0 \pi ^+)
\left( 1-\frac{T(D^+\rightarrow K^0\pi ^+)}
{T(D^+\rightarrow \bar K^0\pi ^+)}\right) \eqno(1)$$
and for the width
$$\Gamma (D^+\rightarrow K_S \pi ^+)=$$
$$=\frac{1}{2}\Gamma (D^+\rightarrow \bar K^0\pi ^+)
\left( 1+2\tan ^2\theta _C Re\hat \rho (D^+\rightarrow K\pi ^+)
+ \tan ^4\theta _C |\hat \rho (D^+\rightarrow K\pi ^+)|^2\right)
$$
$$\simeq \frac{1}{2}\Gamma (D^+\rightarrow \bar K^0\pi ^+)\cdot
\left( 1+2\tan ^2\theta _C Re[\hat \rho (D^+\rightarrow K\pi ^+)]
\right)\, , \eqno(2)$$
where we have defined a `normalized' ratio of transition
amplitudes $\hat \rho$:
$$-\tan ^2\theta _C\cdot \hat \rho (D^+\rightarrow K\pi ^+)
\equiv \frac{T(D^+\rightarrow K^0\pi ^+)}
{T(D^+\rightarrow \bar K^0\pi ^+)}\, . \eqno(3)$$
Likewise for $K_L$ in the final state
$$\Gamma (D^+\rightarrow K_L \pi ^+)\simeq
\frac{1}{2}\Gamma (D^+\rightarrow \bar K^0\pi ^+)\cdot
\left( 1-2\tan ^2\theta _C Re[\hat \rho (D^+\rightarrow K\pi ^+)]
\right) \eqno(4)$$
Therefore
$$\frac{\Gamma (D^+\rightarrow K_S \pi ^+)-
\Gamma (D^+\rightarrow K_L \pi ^+)}
{\Gamma (D^+\rightarrow K_S \pi ^+)+
\Gamma (D^+\rightarrow K_L \pi ^+)}\simeq 2\tan ^2\theta _C \cdot
Re[\hat \rho (D^+\rightarrow K\pi ^+)]\eqno(5)$$
Very roughly one expects $|\hat \rho |\sim {\cal O}(1)$; assuming
factorization (thus eo ipso neglecting final state interactions) and
using the BSW form factors \cite{BSW} one estimates:
$$\hat \rho (D^+\rightarrow K\pi ^+)\simeq -1.2 \eqno(6)$$
Eq.(5) shows that a difference between
$BR(D^+\rightarrow K_S\pi ^+)$ and $BR(D^+\rightarrow K_L\pi ^+)$
is generated by the DCSD mode $D^+\rightarrow K^0\pi ^+$; the
transition amplitude for the latter can thus be extracted once
$D\rightarrow K_L \pi$ has been observed. Numerically one expects that
difference to be around 10 \% or so. Measuring
$D^+\rightarrow K_L\pi ^+$
with such a degree of accuracy of course presents a stiff
experimental challenge.

Likewise one finds:
$$\frac{\Gamma (D^0\rightarrow K_S\pi ^0)-
\Gamma (D^0\rightarrow K_L\pi ^0)}
{\Gamma (D^0\rightarrow K_S\pi ^0)+
\Gamma (D^0\rightarrow K_L\pi ^0)}\simeq
2\tan ^2\theta _C \cdot Re[\hat \rho (D^0\rightarrow K\pi ^0)]
\eqno(7)$$
with $-\tan ^2\theta _C\cdot \hat \rho (D^0\rightarrow K\pi ^0)
\equiv T(D^0\rightarrow K^0\pi ^0)/T(D^0\rightarrow \bar K^0 \pi ^0)$;
factorization yields $\hat \rho (D^0\rightarrow K\pi ^0)\simeq 1$.

\subsection{$D\rightarrow K\pi \pi$}

An analogous analysis can be performed for
$D\rightarrow K\pi \pi$ decays. Yet these three-body final states
exhibit considerably more complexity generating
additional intriguing features: they are fed by several
channels, namely via
$K^*\pi$, $K\rho$ and $(K\pi \pi )_{non-res}$ intermediate states;
furthermore the shape of the Dalitz plot distributions get substantially
distorted by the interference between CAD and DCSD modes involving
neutral kaons -- an effect that has to be incorporated into a detailed
analysis of the Dalitz plot; lastly there are several different charge
combinations possible for the $K\pi \pi$ final states.
\subsubsection{$D^+\rightarrow K\pi \pi$}
There are four different charge combinations, namely
$D^+\rightarrow K^-\pi ^+\pi ^+$,
$D^+\rightarrow K_S\pi ^0\pi ^+$,
$D^+\rightarrow K^+\pi ^-\pi ^+$ and
$D^+\rightarrow K^+\pi ^0\pi ^0$.
The first decay constitutes a pure CAD mode and
the last two pure DCSD channels; the second one represents the
interplay between a CAD and DCSD mode.

The mode $D^+\rightarrow \bar K^{*0}\pi ^+$ can be observed through the two
different final states $D^+\rightarrow [K^-\pi ^+]_{K^*}\pi ^+$ or
$D^+\rightarrow [K_S\pi ^0]_{K^*}\pi ^+$. The former transition is a
pure CAD and thus
$$\Gamma (D^+\rightarrow [K^-\pi ^+]_{K^*}\pi ^+) =
\frac{2}{3}\Gamma (D^+\rightarrow \bar K^{0*}\pi ^+)\, ,\eqno(8)$$
whereas the latter is both a CAD and a DCSD with its rate reflecting
their interference:
$$ \Gamma (D^+\rightarrow [K_S\pi ^0]_{K^*}\pi ^+) \simeq
\frac{1}{6}\Gamma (D^+\rightarrow \bar K^{0*}\pi ^+)
\left( 1+2\tan ^2\theta _C Re[\hat \rho (D^+\rightarrow K^*\pi ^+)]
\right)
\eqno(9)$$
with $\tan ^2\theta _C\cdot \hat \rho (D^+\rightarrow K^*\pi ^+)
\equiv -T(D^+\rightarrow K^{0*}\pi ^+)/
T(D^+\rightarrow \bar K^{0*}\pi ^+)$. Naively one expects
$\Gamma (D^+\rightarrow [K_S\pi ^0]_{K^*}\pi ^+) \simeq
\frac{1}{4}\Gamma (D^+\rightarrow \bar [K^-\pi ^+]_{K^*}\pi ^+)$;
yet the presence of DCSD -- $\hat \rho \neq 0$ -- changes this
relationship. Assuming factorization one estimates
$\hat \rho (D^+\rightarrow K^*\pi ^+)\sim -5$. Adding the
$K_S$ mode to the corresponding $K_L$ mode,
one finds through order $\tan ^2\theta _C$:
$$\Gamma (D^+\rightarrow [K_{S,L}\pi ^0]_{K^*}\pi ^+) \simeq
\frac{1}{2}\Gamma (D^+\rightarrow [K^-\pi ^+]_{K^*}\pi ^+)
\eqno(10)$$
Modes with $K_L$ in the final state represent an experimental challenge;
they may, however, be measured in future dedicated charm experiments.

Comparing $\Gamma (D^+\rightarrow [K_S\pi ^0]_{K^*}\pi ^+)$ with
$\Gamma (D^+\rightarrow [K^-\pi ^+]_{K^*}\pi ^+)$, eqs.(8,9),
will allow us to
extract $Re[\hat \rho (D^+\rightarrow K^*\pi ^+)]$.
The result of such a procedure might not be very precise since
the detection efficiencies for $D^+\rightarrow K_S\pi ^0\pi ^+$ and
$D^+\rightarrow K^-\pi ^+\pi ^+$ are quite different from each other;
in any case it
can then be compared with
the direct measurement of
$\Gamma (D^+\rightarrow [K^+\pi ^-]_{K^*}\pi ^+)$.
Similarly one obtains
$$\Gamma (D^+\rightarrow K_S\rho ^+)=
\frac{1}{2}\Gamma (D^+\rightarrow \bar K^0\rho ^+)\cdot
\left( 1+2\tan ^2\theta _C\, Re[\hat \rho (D^+\rightarrow K\rho ^+)]
\right) \eqno(11)$$
\subsubsection{$D^0\rightarrow K\pi \pi$}
An analogous procedure is applied to $D^0\rightarrow K\pi \pi$ decays.
The final state $K_S \pi ^+\pi ^-$ is reached via
$K^{-*}\pi ^+$, $K^{+*}\pi ^-$,
$K_S \rho ^0$ and $(K_S \pi ^+\pi ^-)_{non-res}$ intermediate states
generated by CAD as well as DCSD; likewise
$K_S\pi ^0 \pi ^0$ is fed by $K_S^*\pi ^0$ and
$(K_S\pi ^0 \pi ^0)_{non-res}$. The mode
$D^0\rightarrow \bar K^{*0}\pi ^0$ can be
observed in the pure CAD $D^0 \rightarrow [K^-\pi ^+]_{K^*}\pi ^0$ and
in the channel $D^0 \rightarrow [K_S\pi ^0]_{K^*}\pi ^0$ to which
CAD and DCSD contribute coherently.
Comparing those two channels allows to extract
$Re[\hat \rho (D^0\rightarrow K^*\pi ^0)]$, which can also
be compared to the direct  measurement of the DCSD rate
in $D^0 \rightarrow [K^+\pi ^-]_{K^*}\pi ^0$.

There emerges a new feature that can help considerably here: a detailed
analysis of the Dalitz plot in $D^0\rightarrow K_S\pi ^+\pi ^-$ allows
us to determine the strength of the DCSD mode
$D^0\rightarrow K^{*+} \pi ^-$ without reference to any other final
state. One best studies the interference region between
$D^0\rightarrow (K_S\pi ^-)_{K^{*-}}\pi ^+$ and
$D^0\rightarrow (K_S\pi ^+)_{K^{*+}}\pi ^-$ rather than the pure
$D^0\rightarrow (K_S\pi ^+)_{K^{*+}}\pi ^-$ region!

\section{Final State Interactions in $D\rightarrow K+\pi 's$ Decays}

As discussed below, the fact that CAD and DCSD can contribute
coherently to $D\rightarrow K_S\pi ,\, K_S\pi \pi$ opens up
the possibility for a CP asymmetry to emerge in these channels.
Such an asymmetry depends on the strong phase shifts in these final
states. Those cannot be predicted with the presently available
theoretical tools -- yet they can be measured.

The presence of DCSD transitions $D^{+,0}\rightarrow
K^0\pi ^{+,0}$ changes the isospin decomposition of the
observed $D\rightarrow K_S \pi$ amplitudes; in particular
$T(D^+\rightarrow K_S\pi ^+)$ is no longer described by a pure
$I=3/2$ amplitude, but contains also $I=1/2$ contributions. One then has
to adopt the following procedure: first measure
$D^+\rightarrow K_S \pi ^+$ vs. $D^+\rightarrow K_L \pi ^+$
and $D^0\rightarrow K_S \pi ^0$ vs. $D^0\rightarrow K_L \pi ^0$
to isolate the amplitudes for $D^+\rightarrow \bar K^0 \pi ^+$
and $D^0\rightarrow \bar K^0 \pi ^0$, respectively. The usual
isospin decomposition is then applied to them together with
$T(D^0\rightarrow K^- \pi ^+)$.

An analogous analysis is applied to $D\rightarrow K\rho$ modes. To
extract the phase shifts rigorously one has to measure
$D^+\rightarrow K_S\rho ^+, \, K_L\rho ^+$,
$D^0 \rightarrow K_S\rho ^0, \, K_L\rho ^0, \,
K^-\rho ^+$.

For the $D\rightarrow K^*\pi$ transitions one has additional handles that
considerably facilitate the analysis: for one can rely on the channels
$D^+\rightarrow \bar K^{0*}\pi ^+$,
$D^0\rightarrow \bar K^{0*}\pi ^0$ and $D^0\rightarrow K^{-*}\pi ^+$,
where the state $\bar K^{0*}$ is identified via the mode
$\bar K^{0*}\rightarrow K^-\pi ^+$; i.e., one does not
have to use final states involving $K_L$ mesons.
This has been discussed above in subsections 1.2.1 and 1.2.2.

\section{Direct CP Asymmetries in $D\rightarrow K_S+\pi 's$}

For a direct CP asymmetry to become observable in a partial decay rate,
two amplitudes with a nontrivial weak as well as nontrivial strong
phase difference --
$\Delta \phi ^{weak}\neq 0 \neq \Delta \alpha ^{strong}$ -- have to
contribute coherently \cite{CARTER}:
$$T(D\rightarrow f)=|M_1|+
e^{i\Delta \alpha ^{strong}}e^{i\Delta \phi ^{weak}}|M_2|$$
$$T(\bar D\rightarrow \bar f)=|M_1|+
e^{i\Delta \alpha ^{strong}}e^{-i\Delta \phi ^{weak}}|M_2|$$
$$\frac{\Gamma (D\rightarrow f)-\Gamma (\bar D\rightarrow \bar f)}
{\Gamma (D\rightarrow f)+\Gamma (\bar D\rightarrow \bar f)}=
\frac{2\rho _M \sin \Delta \alpha ^{strong}\sin \Delta \phi ^{weak}}
{1+\rho _M^2+2\rho _M\cos \Delta \alpha ^{strong}\cos \Delta \phi ^{weak}}
\eqno(12)$$
where $\rho _M=|M_2/M_1|$.

As discussed above two different isospin amplitudes contribute to
$D^+\rightarrow K_S\pi ^+$, $K_S\rho ^+$, $[K_S\pi ^0]_{K^*}\pi ^+$
and $D^0\rightarrow K_S\rho ^0$. Thus one of the two conditions for the
emergence of a direct CP asymmetry is satisfied for these channels, for
we know that the strong phase shifts are large, at least for the
$K\pi$ and $K^*\pi$ final states. The question remaining open is then
the size of $\Delta \phi ^{weak}$.

While the KM ansatz provides a weak phase in charm decays,
it is of order $sin^4\theta _C \sim 10^{-3}$ and thus of only
academic interest in $D\rightarrow K_S +\pi 's$ decays, since it would lead
to a CP asymmetry of order
$2\tan ^2\theta _C\cdot |\hat \rho|\cdot \sin ^4\theta _C\sim
{\cal O}(10^{-4})$ only \cite{BS,BTAUCHARM,GOLDEN,BUCELLA}.
Yet New Physics could intervene -- in particular through the
DCSD transitions -- to generate a much more sizeable
weak phase between these two
amplitudes allowing for a direct CP asymmetry to reach the
per cent level!

Such asymmetries would have four important features:

\noindent $\bullet$ As discussed in the preceding section, the strong
phase shifts relevant for the observable size of the asymmetry can
be determined experimentally independant of CP violation. For
example, in $D^{\pm}\rightarrow K^*\pi ^{\pm}$ decays one has six
modes, namely (i) $D^+\rightarrow [K^-\pi ^+]_{K^*}\pi ^+$,
(ii) $D^-\rightarrow [K^+\pi ^-]_{K^*}\pi ^-$,
(iii) $D^+\rightarrow [K^+\pi ^-]_{K^*}\pi ^+$,
(iv) $D^-\rightarrow [K^-\pi ^+]_{K^*}\pi ^-$,
(v) $D^+\rightarrow [K_S\pi ^0]_{K^*}\pi ^+$ and
(vi) $D^-\rightarrow [K_S\pi ^0]_{K^*}\pi ^-$. Reactions (i) and (ii)
[(iii) and (iv)] are pure CAD [DCSD] and
therefore have to have equal rates.
Reactions (v) and (vi), on the other hand, contain interference between
CAD and DCSD and thus can exhibit a difference.
Therefore there are four
independant observables -- $\Gamma (D^+\rightarrow [K^-\pi ^+]_{K^*}\pi ^+)$,
$\Gamma (D^+\rightarrow [K^+\pi ^-]_{K^*}\pi ^+)$ and
$\Gamma (D^{\pm}\rightarrow [K_S\pi ^0]_{K^*}\pi ^{\pm})$ -- that allow to
extract
separately the four unknowns, namely the two real amplitudes $|M_1|$, $|M_2|$
and the phase differences $\Delta \alpha ^{strong}$ and
$\Delta \phi ^{weak}$. This can be discussed in close analogy to the
more familiar case of the CP asymmetry in $B^{\pm}\rightarrow DK^{\pm}$
decays \cite{WYLER,PAIS}.

\noindent $\bullet$ New Physics sources for CP violation can quite
conceivably and naturally lead to CP asymmetries that are larger by two
orders of magnitude! For example, New Physics could generate an
imaginary part in the DCSD amplitude of relative strength, say, 5\%;
this would lead to an asymmetry $\sim 0.5$\%.

\noindent $\bullet$ Since
$\Gamma (D^+\rightarrow K_{S,L}\pi ^+)=
\Gamma (D^-\rightarrow K_{S,L}\pi ^-)$ has to hold, a difference in
$\Gamma (D^+\rightarrow K_S\pi ^+)$ vs.
$\Gamma (D^-\rightarrow K_S\pi ^-)$ has to be compensated by a
difference in $\Gamma (D^+\rightarrow K_L\pi ^+)$ vs.
$\Gamma (D^-\rightarrow K_L\pi ^-)$ that is equal in size and
opposite in sign!

\vspace{1cm}

\noindent {\bf Acknowledgements}

\noindent This work was supported by the
National Science Foundation under grant number PHY 92-13313
and by the Dept. of Energy under grant number
DE-FG02-91ER40654.


\begin{thebibliography}{99}

\bibitem {BSW}
M. Bauer, B. Stech and M. Wirbel, Z. Phys. {\bf C23}, 103 (1987).

\bibitem{CARTER}
A. Carter, A. Sanda, Phys. Rev. {\bf D23}, 1567 (1981).

\bibitem{BS}
I.I. Bigi, A.I. Sanda, Phys. Lett. {\bf B171}, 320 (1986).

\bibitem{BTAUCHARM}
I.I. Bigi, in: Proceedings of the Tau-Charm Factory Workshop,
Stanford, CA, May 1989, ed. by L.V. Beers,
SLAC Report {\bf 343}, p. 169.

\bibitem{GOLDEN}
M. Golden, B. Grinstein, Phys. Lett. {\bf B222}, 501 (1989).

\bibitem{BUCELLA}
F. Buccella {\it et al.}, Phys. Lett. {\bf B302}, 319 (1993).

\bibitem{WYLER}
M. Gronau, D. Wyler, Phys. Lett. {\bf 265}, 172 (1991).

\bibitem{PAIS}
I.I. Bigi, A.I. Sanda, Phys. Lett. {\bf B211}, 213 (1988).


\end{thebibliography}
\end{document}